# A Classification Scheme for Inverse Design of Molecules: from Targeted Electronic Properties to Atomicity

Alain B. Tchagang and Julio J. Valdés

National Research Council Canada
Digital Technologies Research Centre
M-50, 1200 Montréal Road Ottawa ON K4A 0S2, Canada
`{alain.tchagang, julio.valdes}@nrc-crnc.gc.ca`

**Abstract.** In machine learning and molecular design, there exist two approaches: discriminative and generative. In the discriminative approach dubbed forward design, the goal is to map a set of features/molecules to their respective electronics properties. In the generative approach dubbed inverse design, a set of electronics properties is given and the goal is to find the features/molecules that have these properties. These tasks are very challenging because the chemical compound space is very large. In this study, we explore a new scheme for the inverse design of molecules based on a classification paradigm that takes as input the targeted electronic properties and output the atomic composition of the molecules (i.e. atomicity or atom counts of each type in a molecule). To test this new hypothesis, we analyzed the quantum mechanics QM7b dataset consisting of 7211 small organic molecules and 14 electronic properties. Results obtained using twenty three different classification approaches including a regularized Bayesian neural network show that it is possible to achieve detection/prediction accuracy > 90%.

## 1 Introduction

The objective of molecular design (MD) for new drugs and new materials is to be able to generate new molecules with targeted properties. This task is very challenging because the chemical compound space (CCS) is very large, and traditional approaches which are based on trial-by-errors are very expensive and time consuming [1]. Recent technological advances have shown that data-to-knowledge approaches are beginning to show enormous promise within MD. Intelligent exploration and exploitation of the vast molecular property space has the potential to alleviate the cost, risks, and time involved in trial-by-error approach experiment cycles used by current techniques to identify useful compounds [1].

In machine learning (ML) and MD, there exist two approaches: discriminative and generative models. In the discriminative approach dubbed forward design, the goal is to map a set of features/molecules to their respective electronics properties. In the generative approach dubbed inverse design, a set of targeted electronics properties is given and the goal is to find the features/molecules that have these properties [2].

In this study, we focus on the inverse design. Generative models based on deep learning techniques have recently been widely used to tackle this problem and may offer interesting and efficient solutions to MD. In [3] the authors used a variational autoencoder (VAE) to optimize the molecular properties in a hidden space, where molecules are expressed as real vectors. They applied their technique to improve the partition of drug and the emission rate of organic light emitting diode candidates. In [4], an adversarial autoencoder (AAE) and a Bayesian optimization approach were combined and used to generate ligands specific to the dopamine type 2 receptors. The authors of [5] compared the VAE and AAE as a molecular generation model in terms of the reconstruction error and variability of molecular fingerprints. In addition to the above approaches based on autoencoders, generative adversarial networks known as GAN [6], transfer learning [7] and reinforcement learning methods [8, 9] have also been explored in the context of inverse MD. A good mini-review on this topic is available in [10].

In this paper, we follow a more direct approach. We explore a ML based classification scheme for the inverse MD. Given a set of targeted electronic properties, we show that it is possible to infer the atomicity or atomic composition of the molecules (i.e. atom counts of each type in a molecule) that correspond to the given set of properties. To test this new hypothesis, the classification scheme is used to analyze the publicly available QM7b dataset [11, 12]. Analysis results using twenty tree different ML classification techniques including a Bayesian regularized neural networks (BRNN) gave detection/prediction accuracy $> 90\%$.

The rest of this paper is organized as follows. In Section II, the dataset used in this study is described. Section III provides a detailed description of the proposed method. Section IV presents the results and discussions. Section V concludes this study.

## 2    QM7b Dataset

The QM7b dataset used in this study is an extension of the QM7 dataset [11]. It is made of 7211 small organic molecules, each molecule is composed of one of the following six atoms: Carbon (C), Chlorine (Cl), Hydrogen (H), Azote (N), Oxygen (O), and Sulfur (S). 14 electronic properties for each of the 7211 molecules are available and computed using QM first principle [2]. These 14 properties make the QM7b dataset interesting for multitasking learning and they correspond to: PBE0 atomization energies, zindo-excitation-energy-with-the-m, zindo-highest-absorption intensity, zindo-homo, zindo-lumo, zindo-1st-excitation-energy, zindo-ionization-potential, zindo-electron-affinity, PBE0-homo, PBE0-lumo, GW-homo, GW-lumo, PBE0 polarizability, and SCS polarizability. More details relative to this dataset can be obtained in [12, 13].

## 3 Methods

Our starting point is the atomic composition (AC) of molecules. Next a classifier is design, with the goal of taking a set of molecular electronic properties as input and to generate the AC of the molecules that correspond to such properties.

### 3.1 Atomicity, Atom Counts or Atomic Composition (AC)

Let's define $\Omega = \{\Omega_1, \Omega_2, \ldots, \Omega_m, \ldots, \Omega_M\}$, the set of possible molecules in the chemical compound space (CCS). By construction, this space is very large. In this study, we will assume that it is bounded by M. Let's define A the set of unique atoms that make $\Omega$. A is bounded by K and it is defined as: $A = \{A^1, A^2, \ldots, A^k, \ldots, A^K\}$. Let's suppose that there exists a chemical operator that combines atoms among them in a specific numbers $u_{mk}$ and according to the laws of chemistry to form a stable molecule $\Omega_m$. The chemical formulae of $\Omega_m$ can be written as: $\Omega_m \equiv u_{m1}A^1 u_{m2}A^2 \ldots u_{mk}A^k \ldots u_{mK}A^K$, or as in chemical textbook.

$$\Omega_m \equiv A^1_{u_{m1}} A^2_{u_{m2}} \ldots A^k_{u_{mk}} \ldots A^K_{u_{mK}} \tag{1}$$

The AC of molecule $\Omega_m$ in the atomic space $[A^1 \ A^2 \ \ldots \ A^k \ \ldots \ A^K]$ is defined as $[u_{m1} \ u_{m2} \ \ldots \ u_{mk} \ \ldots \ u_{mK}]$, where $u_{mk}$ is a positive integer that represents the number of atom $A^k$ in molecule $\Omega_m$. The AC of the M molecules in the atomic space $[A^1 \ A^2 \ \ldots \ A^k \ \ldots \ A^K]$ can be organized in an M×K matrix U, **Equation 2**.

$$U = \begin{bmatrix} \Omega_1 \\ \Omega_2 \\ \vdots \\ \Omega_m \\ \vdots \\ \Omega_M \end{bmatrix} = \begin{bmatrix} u_{11} & u_{12} & \ldots & u_{1k} & \ldots & u_{1K} \\ u_{21} & u_{22} & \ldots & u_{2k} & \ldots & u_{2K} \\ \vdots & \vdots & \ldots & \vdots & \ldots & \vdots \\ u_{m1} & u_{m2} & \ldots & u_{mk} & \ldots & u_{mK} \\ \vdots & \ldots & \ldots & \vdots & \ldots & \vdots \\ u_{M1} & u_{M2} & \ldots & u_{Mk} & \ldots & u_{MK} \end{bmatrix} \tag{2}$$

Row u(m,:) of U corresponds to the AC of the m$^{th}$ molecule ($\Omega_m$) in the atomic space $[A^1 \ A^2 \ \ldots \ A^k \ \ldots \ A^K]$. Column u(:,k) corresponds to the number of atom $A^k$ in each molecule in $\Omega$. For example, given a set of seven molecules: $\Omega = \{CH_4, C_2H_2, C_3H_6, C_2NH_3, OC_2H_2, ONC_3H_3, SC_3NH_3\}$. The set of unique atoms that makes $\Omega$ is A = {C, H, N, O, S} and K = 5. The matrix U is given by:

$$U = \begin{bmatrix} CH_4 \\ C_2H_2 \\ C_3H_6 \\ C_2NH_3 \\ OC_2H_2 \\ ONC_3H_3 \\ SC_3NH_3 \end{bmatrix} = \begin{bmatrix} 1 & 4 & 0 & 0 & 0 \\ 2 & 2 & 0 & 0 & 0 \\ 3 & 6 & 0 & 0 & 0 \\ 2 & 3 & 1 & 0 & 0 \\ 2 & 2 & 0 & 1 & 0 \\ 3 & 3 & 1 & 1 & 0 \\ 3 & 3 & 1 & 0 & 1 \end{bmatrix} \quad (3)$$

It is obvious that this representation is not unique. That is two molecules with identical AC may have different electronic properties. Isomers are great examples in this case. They are compound with the same molecular formulas but that are structurally different in some way, and they can have different chemical, physical and biological properties [14]. It is also worth to note that such molecular representation had been explored in the past in quantitative structure activity relationship and correspond to a different form of the Atomistic index developed by Burden [15].

### 3.2  Electronic Properties of Molecules

The electronic properties of the M molecules can also be organized in an M×L matrix, where L is the number of electronic properties.

$$p = \begin{bmatrix} p_{11} & p_{12} & \ldots & p_{1l} & \ldots & p_{1L} \\ p_{21} & p_{22} & \ldots & p_{2l} & \ldots & p_{2L} \\ \vdots & \vdots & \ldots & \vdots & \ldots & \vdots \\ p_{m1} & p_{m2} & \ldots & p_{ml} & \ldots & p_{mL} \\ \vdots & \ldots & \ldots & \vdots & \ldots & \vdots \\ p_{M1} & p_{M2} & \ldots & p_{Ml} & \ldots & p_{ML} \end{bmatrix} \quad (4)$$

The $m^{th}$ row of p represents the electronic properties of the $m^{th}$ molecule, $p_{ml}$ is a real number and it corresponds to the $l^{th}$ electronic property of the $m^{th}$ molecule. These properties are obtained from computational QM first principles [2, 12].

### 3.3  From Electronic Properties to Atomic Composition of Molecules

Given the set of electronic properties of a molecule, is it possible to infer its atomic composition uniquely from the given set of properties? To answer this question, we adopt a ML based classification scheme. In ML, classification is a type of supervised learning where a training set of correctly identified observations (electronic properties in this study) and classes (atomic composition in this study) is available. The goal is to assign new observations to a class. Several classifiers have been developed in the literature: support vector classifiers, K-nearest neighbor classifiers, classification tree, discriminant classifiers and classifiers based on ensemble methods [16].

Neural network methods are also widely used to solve classification problems [16]. Multilayer feed-forward networks are the most popular and a large number of training algorithms have been proposed. Compared to other non-linear techniques, in multi-layer NNs, the measure of similarity is learned essentially from data and implicitly given by the mapping onto increasingly many layers. In general, NNs are more flexible and make fewer assumptions about the data. However, it comes at the cost of being more difficult to train and regularize [12]. In this paper, we used the Bayesian regularization neural network (BRNN) method [15, 17, 18, 19, 20].

Bayesian methods are optimal methods for solving learning problems. Any other method not approximating them should not perform as well on average. They are very useful for comparison of data models as they automatically and quantitatively embody "Occam's Razor" [21]. Complex models are automatically self-penalizing under Bayes' Rule. Bayesian methods are complementary to NNs as they overcome the tendency of an over flexible network to discover nonexistent, or overly complex, data models. Unlike a standard back-propagation NN training method where a single set of parameters (weights, biases, etc.) are used, the Bayesian approach to NN modeling considers all possible values of network parameters weighted by the probability of each set of weights. Bayesian inference is used to determine the posterior probability distribution of weights and related properties from a prior probability distribution according to updates provided by the training set $D$ using the BRNN model, $H_i$. Where orthodox statistics provide several models with several different criteria for deciding which model is best, Bayesian statistics only offer one answer to a well-posed problem.

$$P(w|D,H_i) = \frac{P(D|w,H_i)P(w|H_i)}{P(D|H_i)} \qquad (5)$$

Bayesian methods can simultaneously optimize the regularization constants in NNs, a process which is very laborious using cross-validation [17].

## 4  Results and Discussions

As we mentioned earlier, the QM7b dataset used in this study is the one published in [12] and available from the quantum-machine.org website. It is composed of M = 7211 molecules and contains up to six types of atoms: Carbon (C), Chlorine (Cl), Hydrogen (H), Oxygen (O), Azote (N), and Sulfur (S). The set of unique atoms is A = {C, Cl, H, N, O, S}. The matrix U of AC is of size M×K = 7211×6. The largest molecule is made of 23 atoms. The p matrix of electronic properties is of size 7211×14. The 7211 molecules are unique and the set does not include isomers. We used a set of twenty two classifiers implemented within the Matlab R2016b environment, and the neural network with Bayesian regularization as also implemented in the Matlab R2016b environment. First 5-fold cross validation is used across all the classifiers and the BRNNs. Furthermore, the QM7b dataset is randomly divided into 70% training and 15% validation and 15% testing sets and only applied to the neural network. This second step was performed to evaluate the BRNN classifier on data not previously seen by the classifier.

### 4.1 Results

**Figure 1** shows the architecture of the model. It takes as input the 14 electronic properties and generates the atom counts of each type that makes the given molecule.

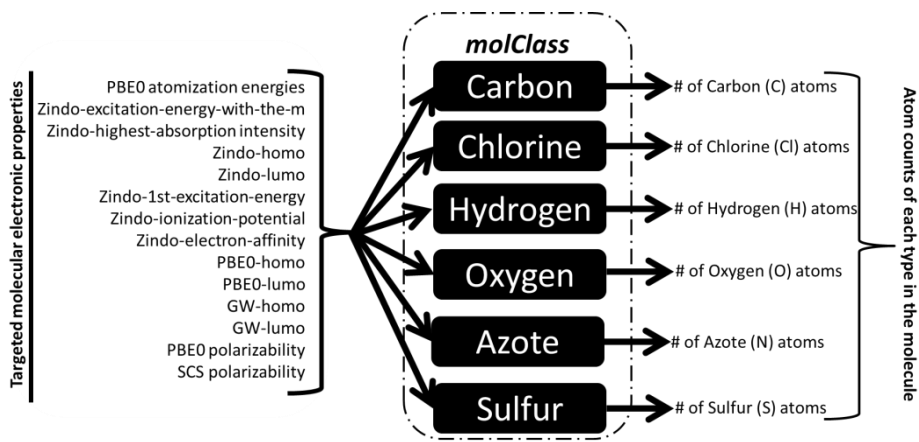

**Fig. 1.** Architecture of the classifier

From the AC matrix (i.e. U), we infer that the Carbon has seven classes: {1, 2, 3, 4, 5, 6, 7}. That is, each molecule in the QM7b can have one or up to seven atoms of Carbon. The Chlorine has three classes: {0, 1, 2}. The Hydrogen has seventeen classes: {0, 1, 2, 3, 4, 5, 6, 7, 8, 9, 10, 11, 12, 13, 14, 15, 16}. The Azote has four classes: {0, 1, 2, 3}. The Oxygen has four classes: {0, 1, 2, 3}, and the Sulfur has two classes: {0, 1}.

**Table 1** shows the classification results of the twenty two classical classifiers. On the detection/prediction of atom counts of Carbon, Chlorine, Hydrogen, Azote, Oxygen and Sulfur in each molecule, Cubic SVM (89.7%), Cubic SVM (99.8%), Ensemble Bagged Trees (86.4%), Ensemble Subspace KNN (85.8), Ensemble Subspace KNN (90.3%), and Quadratic SVM (99.7%) achieved the best results respectively. In general and across the six atom classes, Cubic SVM, Ensemble Subspace KNN, and Ensemble Bagged Trees gave the best classification accuracy results ($> 85\%$).

In the literature, there is no clear and rational approach on how to select the number of neurons and hidden layers of a NN. The middle ground is usually to select an architecture that will neither under-fit nor over-fit the network. In this study we tested several architecture based on some empirical observations also coming from the literature with the goal for avoiding under-fitting and overfitting of the model. **Table 2** shows the results obtained using different NN architectures for 5-fold cross validation. **Table 3**, the results with 70%, 15% and 15% training, validation and testing.

**Table 1**: Classification accuracy (%) using 22 classifiers with 5-fold cross validation

| Classifiers | Cabon | Chlorine | Hydrogen | Azote | Oxygen | Sulfur |
|---|---|---|---|---|---|---|
| Complex tree | 72.1 | 99.4 | 68.7 | 69.2 | 78.8 | 97.9 |
| Medium tree | 64.6 | 99.4 | 58.2 | 62.9 | 68.3 | 97.7 |
| Simple tree | 56.8 | 99.4 | 46.6 | 51.3 | 60.0 | 97.1 |
| Linear Discriminant | 54.0 | 97.3 | 45.8 | 45.2 | 58.3 | 89.6 |
| Quadratic Discriminant | Failed | Failed | Failed | 47.9 | 59.0 | 87.4 |
| Linear SVM | 71.2 | 99.5 | 65.4 | 57.5 | 67.0 | 99.4 |
| Quadratic SVM | 85.6 | 99.7 | 80.8 | 80.8 | 77.8 | **99.7** |
| Cubic SVM | **89.7** | **99.8** | 85.7 | 85.6 | 83.1 | 99.6 |
| Fine Gaussian SVM | 82.4 | 98.5 | 76.0 | 81.8 | 79.3 | 98.2 |
| Medium Gaussian SVM | 82.6 | 99.5 | 79.0 | 76.7 | 75.0 | 99.4 |
| Coarse Gaussian SVM | 69.9 | 99.5 | 61.2 | 59.4 | 63.4 | 96.6 |
| Fine KNN | 84.9 | 99.7 | 77.5 | 82.3 | 81.9 | 99.2 |
| Medium KNN | 79.7 | 99.6 | 70.3 | 76.8 | 75.2 | 98.5 |
| Coarse KNN | 69.3 | 99.5 | 54.1 | 65.0 | 65.6 | 96.1 |
| Cosine KNN | 78.8 | 99.6 | 68.0 | 75.1 | 75.4 | 98.4 |
| Cubic KNN | 78.5 | 99.6 | 68.5 | 74.9 | 72.9 | 98.3 |
| Weighted KNN | 84.2 | 99.7 | 76.5 | 81.1 | 80.2 | 98.9 |
| Ensemble Boosted Trees | 71.2 | 99.5 | 60.4 | 67.6 | 74.0 | 99.3 |
| Ensemble Bagged Trees | 88.0 | 99.6 | **86.4** | 85.4 | 88.4 | 99.2 |
| Ensemble Subspace Discriminant | 65.6 | 99.5 | 53.2 | 52.5 | 62.0 | 97.1 |
| Ensemble Subspace KNN | 86.1 | 99.7 | 82.8 | **85.8** | **90.3** | 98.5 |
| Ensemble RUSBoosted Trees | 50.7 | 90.1 | 52.1 | 58.4 | 64.7 | 97.2 |

**Table 2**: Classification accuracy (%) using different NN architectures, with 5-fold cross validation

| Network architecture | Carbon | Chlorine | Hydrogen | Azote | Oxygen | Sulfur |
|---|---|---|---|---|---|---|
| [2] | 77.8 | 100 | 69.6 | 72.9 | 72.0 | 100 |
| [14 2] | 93.2 | 100 | 83.2 | 91.3 | 87.5 | 100 |
| [3] | 82.4 | 100 | 70.2 | 75.5 | 75.9 | 100 |
| [14 3] | 90.3 | 100 | 83.5 | 89.3 | 87.9 | 100 |
| [4] | 84.5 | 100 | 74.1 | 77.0 | 76.8 | 100 |
| [14 4] | 92.3 | 100 | 82.2 | 91.6 | 87.9 | 100 |
| [7] | 88.1 | 100 | 80.1 | 83.9 | 82.8 | 100 |
| [14 7] | 94.5 | 100 | 88.3 | 93.1 | 90.9 | 100 |
| [17] | 93.9 | 100 | 88.5 | 91.2 | 89.4 | 100 |
| [14 17] | **95.9** | 100 | **91.2** | **95.3** | **92.1** | 100 |

**Table 3**: Classification accuracy (%) using different NN architectures, using 70% for training and 15% for validation and 15% for testing

| Network architecture | Carbon | Chlorine | Hydrogen | Azote | Oxygen | Sulfur |
|---|---|---|---|---|---|---|
| [2] | 76.9 | 99.9 | 68.5 | 72.0 | 71.0 | 99.8 |
| [14 2] | 92.5 | 99.9 | 82.3 | 90.2 | 86.7 | 100 |
| [3] | 81.2 | 99.9 | 69.4 | 74.9 | 74.6 | 99.9 |
| [14 3] | 89.0 | 100 | 82.6 | 88.1 | 87.2 | 99.9 |
| [4] | 83.4 | 99.9 | 73.4 | 76.5 | 76.3 | 99.9 |
| [14 4] | 91.4 | **100** | 81.0 | 90.5 | 87.6 | 99.9 |
| [7] | 87.7 | **100** | 79.2 | 83.4 | 82.1 | 100 |
| [14 7] | 94.1 | 99.9 | 87.9 | 91.8 | 90.4 | 100 |
| [17] | 93.8 | 99.5 | 87.7 | 90.6 | 88.3 | 99.9 |
| [14 17] | **94.6** | **100** | **90.4** | **94.7** | **91.9** | 99.9 |

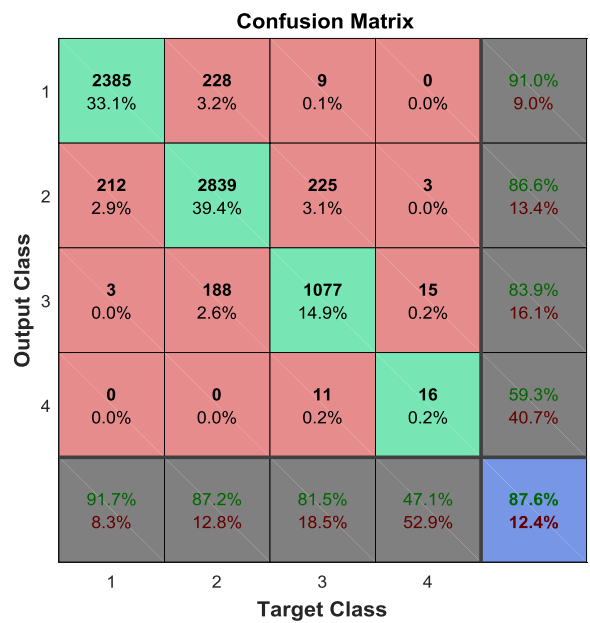

**Fig. 2.** Confusion matrix for detection of the 4 classes of Oxygen with the [14 4] NN architecture. Diagonal shows the number of correct classifications, off diagonal misclassifications.

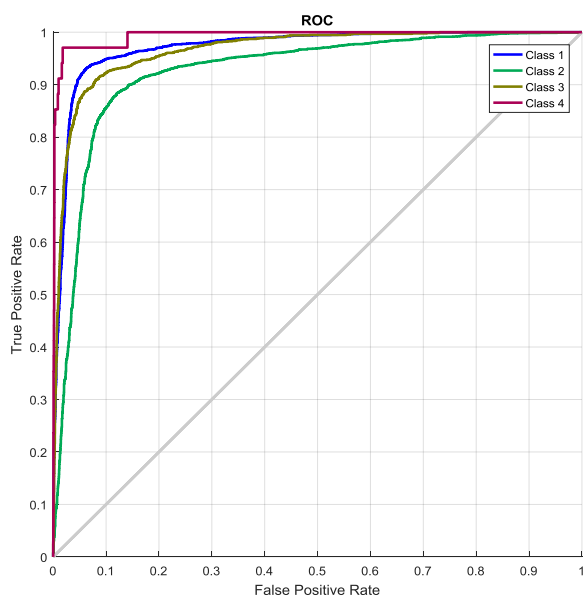

**Fig. 3.** ROC for detection of the 4 classes of Oxygen with the [14 4] NN architecture. Class1, class2, class3, and class4 correspond to molecules with 0, 1, 2, and 3 atoms of Oxygen.

In all, our study suggests that for each atom, the optimal architecture is the one with two hidden layers, with the first hidden layer having the same number as the number of properties (i.e. 14 in this study) and the second hidden layer having the same number of classes as that of the atoms to be predicted (i.e. 7, 3, 17, 4, 4, and 2 for Carbon, Chlorine, Hydrogen, Azote, Oxygen, and Sulfur respectively. The number of Epoch was fixed to 100 but can always be increased to improve the accuracy of the results. All the other Bayesian regularization parameters were set to default as implemented in Matlab R2016 environment. In general the BRNNs approach gave better results compared to the twenty two other classical classifiers. Bayesian regularization also performed better compared to other regularization methods (results not shown here). **Figure 2** and **Figure 3** show the confusion matrix and the ROC for the detection of the number of Oxygen atoms per molecule using the [14 4] NN architecture. Class 1, class 2, class 3, and class 4 correspond to molecules with 0, 1, 2, and 3 atoms of Oxygen respectively. The diagonal of the confusion matrix shows the number of molecules with atom counts that were correctly classified whereas the off-diagonal shows misclassifications on the testing set.

### 4.2    Statistical Significance of Atomic Composition

To test the statistical significance of the AC, random U matrices were generated and used as output to the classification scheme. Results obtained were meaningless and corresponded to noise. The difference between the real AC matrix U and the randomly generated ones show that there exists some interesting information encoded in the AC and represent good features for inverse design of molecules. On the other hand, multiplication of one or more columns of the AC matrix with the same constant left the classification/prediction results unchanged.

### 4.3    Discussions

AC, i.e. atom counts of each type in a molecule is a representation that does not contain any molecular structural information. But our analysis suggests a correlation between the AC and the electronic properties of molecules. One of the most interesting results to mention here is that, the Hydrogen atoms, one of the ingredients of the molecules under investigation, with 17 classes could be detected with accuracy of 86.4% and 90.4% by Ensemble Bagged Trees and BRNNs respectively. Furthermore Sulfur with only 2 classes was detected with accuracy close to ~100% by the majority of the classifiers. These observations contribute to reinforce the correlation-relationship between the electronics properties of molecules and their AC whether the number of classes is higher and more complex or lower and less complex.

The problem we tackle in this study is somehow similar to the well-known iris flower dataset [22] which is widely used in ML. The iris dataset consists of 50 samples from each of three species of Iris (*Iris setosa*, *Iris virginica* and *Iris versicolor*), and consists of four features for each sample: the length and the width of the sepals and petals. Based on the combination of these four features, Fisher developed a linear discriminant model to distinguish the species from each other. In other terms, in the iris dataset, the physical properties of flowers are given and the goal is to use these properties to infer the type of flowers. Similarly, in this study, the electronic properties of molecules are known and the goal is to infer the AC of the molecules corresponding to these electronic properties. But unlike the Iris dataset, we go one step deeper in this study. We not only detect the molecules, we also detect the ingredients that make these molecules (i.e. atom counts of each type in the molecule). From the results obtained there exists a clear correlation between the set of electronic properties and the AC.

As many other ML models, the disadvantage of the classification scheme based inverse design that we developed in this study is limited by the set of domain knowledge used to design the classifier. For example, given that the atom of Carbon only has seven classes: {1, 2, 3, 4, 5, 6, 7} in this study, we can only use it to predict new molecules with only 1 or up to 7 atoms of Carbon. To infer a new molecule with 8 atoms of Carbon for example, the model could be either retrained by integrating knowledge relative to molecules with 8 atoms or Carbon (this involve more data collection), or the model could be extrapolated outside the boundaries of the current classifier. The latter view is a much difficult ML and statistical problem and will be explored in future works.

## 5      Conclusions

Predicting molecular electronic properties quickly and accurately across the chemical compound space is an important problem as the quantum mechanics calculations are typically time consuming and do not scale well to more complex molecules. Machine learning is a natural candidate for solving this problem as it encourages computational units to focus on solving the problem of interest rather than solving the more general Schrödinger equations. In this study, we proposed and validated a machine learning based classification scheme, useful for the inverse design of molecules. Using the QM7b dataset as a testbed, we showed that it is indeed possible to map the electronic properties of molecules to the number of atoms of each type that make these molecules. Our study suggests important results and open new venues for future research in the inverse design of molecules, new drugs and new materials.